\documentclass{moriond}

\def\be{\begin{equation}}
\def\ee{\end{equation}}
\def\bea{\begin{eqnarray}}
\def\eea{\end{eqnarray}}

\usepackage{amsmath}
\usepackage{hyperref}

\usepackage{xspace}

\newcommand{\zeff}{z_{\rm eff}\xspace}
\newcommand{\planck}{\textsc{Planck-2018}\xspace}
\newcommand{\zph}{z_{\rm ph}\xspace}

\usepackage{float}

\newcommand{\Photo}{\vspace{0.2cm}\includegraphics[height=35mm]{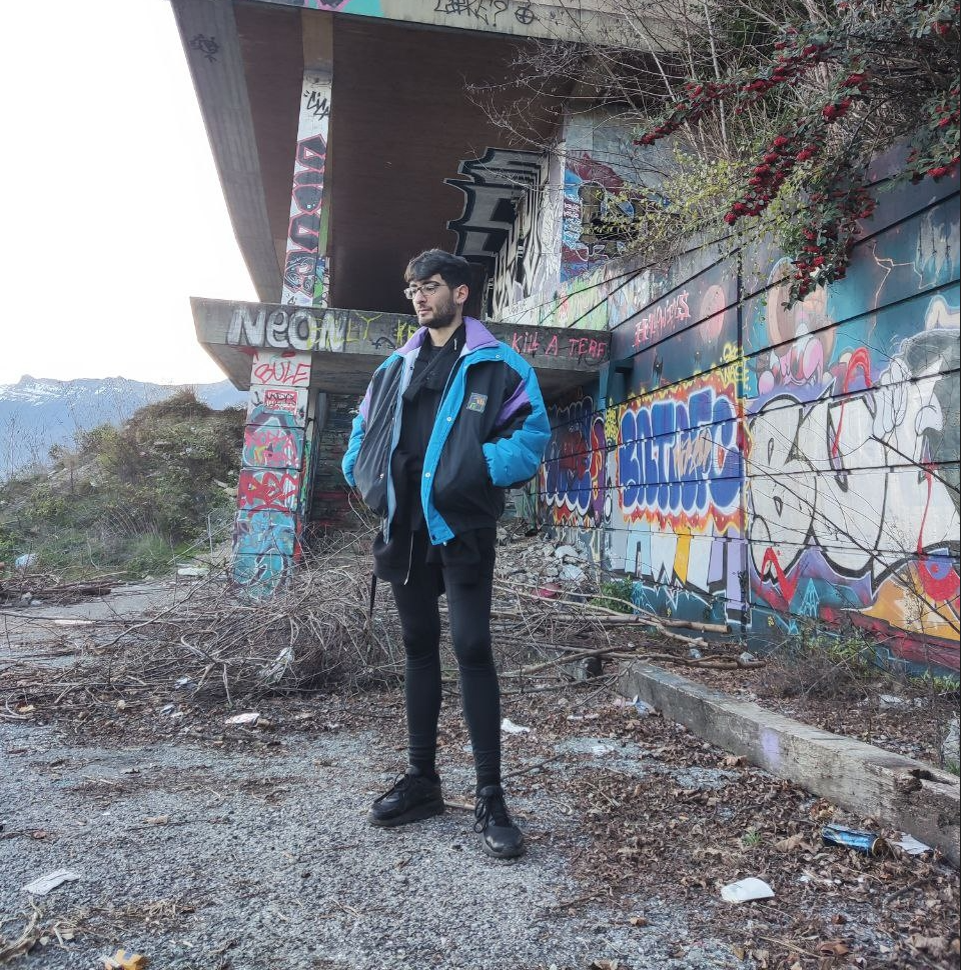}\vspace{-0.15cm}}

\begin{document}
\vspace*{4cm}
\title{Dark Energy Survey: 2.1\% measurement of the Baryon Acoustic Oscillation scale from the final dataset
}

\author{Juan Mena-Fern\'andez on behalf of the Dark Energy Survey Collaboration}

\address{LPSC Grenoble - 53, Avenue des Martyrs 38026 Grenoble, France}

\maketitle\abstracts{
Here, we present the angular diameter distance measurement obtained from the measurement of the Baryonic Acoustic Oscillation (BAO) feature using the completed Dark Energy Survey (DES) data, summarizing the main results of \cite{BAOkey} and \cite{mena2024dark}. We use a galaxy sample optimized for BAO science in the redshift range $0.6<z<1.2$, with an effective redshift of $\zeff=0.85$. Our consensus measurement constrains the ratio of the angular distance to the sound horizon scale to $D_M(\zeff)/r_d = 19.51 \pm 0.41$. This measurement is found to be $2.13\sigma$ below the angular BAO scale predicted by \planck. To date, it represents the most precise measurement from purely photometric data, and the most precise from any Stage-III experiment at such high redshift. The analysis was performed {\it blinded} to the BAO position and is shown to be robust against analysis choices, data removal, redshift calibrations and observational systematics.
}

\section{Introduction}

The Dark Energy Survey\footnote{\url{https://www.darkenergysurvey.org/}} (DES) is a Stage-III photometric galaxy survey designed to constrain the properties of dark energy and other cosmological parameters from multiple probes. DES has performed state-of-the-art analyses of weak gravitational lensing, galaxy clustering and galaxy cluster counts. These probes have also been combined with external cosmic microwave background data. The DES Supernova (SN) program has also broken new grounds in constraining cosmology from $\sim$1,500 type Ia SN \cite{SNkey}. In addition to that, the large data sets and catalogs produced by DES represent a unique source for other cosmological and astronomical analyses.

In this work, we use the complete DES data set, which includes 6 years of observations (2013-2019), to constrain the angular BAO distance scale, summarizing the main outcomes of \cite{BAOkey}. We follow a similar methodology to previous DES analyses, with three main changes. First, in \cite{mena2024dark} we re-optimize the galaxy sample and extend it up to redshift 1.2. Second, we reinforce the redshift validation, considering several independent calibrations and quantifying its possible impact on the BAO measurement. Third, we provide BAO measurements from three types of two-point clustering statistics: angular correlation function (ACF), angular power spectrum (APS) and projected correlation function (PCF). Our reported consensus result comes from the statistical combination of those three measurements, following a strict {\it blinding} policy.

\section{The BAO-Optimized Sample}

The galaxy sample used to measure the BAO signal is a subset of the DES data set selected using the $griz$ bands and a photometric redshift estimate, $\zph$. The selection cuts applied are
\begin{eqnarray}
    1.7<i-z+2(r-i)\ & \text{(color selection)},\\
    17.5<i<19.64+2.894\zph\ & \text{(flux selection)},\label{eq:flux_selection}\\
    i<22.5\ & (i-\text{mag limit}),\\
    0.6<\zph<1.2\ & \text{(photo-$z$ range)}.
\end{eqnarray}
The sample has 15.93 million galaxies, its angular mask covers 4,273 deg${}^2$, it is divided into 6 redshift bins with $\Delta\zph=0.1$ and its effective redshift is $\zeff=0.85$. The cut specified by Eq. \eqref{eq:flux_selection} was optimized using a Fisher forecast algorithm, as detailed in \cite{mena2024dark}.

\section{Methodology}

\subsection{Simulations}

We created a set of 1,952 mock catalogs based on ICE-COLA fast simulations \cite{ferrero2021dark}. They reproduce with high accuracy the main properties of the data: observational volume, abundance of galaxies, redshift distributions, photo-$z$ errors and clustering as a function of redshift. Mock catalogs are key to validating the modeling and quantifying how likely some features we find in the data are.

\subsection{The BAO Fit}

Our approach to measuring the BAO distance scale is based on a template fitting method. We start from the linear power spectrum generated using {\tt CAMB}, $P_{\rm lin}(k)$, and then isolate the {\it no-wiggle} component, $P_{nw}(k)$, as
\begin{equation}
    P(k,\mu) = (b+\mu^2 f)^2\left[ (P_{\rm lin}-P_{\rm nw})e^{-k^2\Sigma^2}+P_{\rm nw}\right],
    \label{eq:pkmu}
\end{equation}
where $b$ is the linear galaxy bias, $\mu^2f$ accounts for redshift-space distortions, and $\Sigma$ models the broadening of the BAO peak due to non-linearities. From this $P(k,\mu)$, we compute the theoretical template, $T_{\rm BAO}(x)$, for our three estimators (ACF, APS and PCF). For the template, we use \planck \cite{Planck} as the reference cosmology and assume $\Lambda$CDM. The model fitted to the data is given by
\begin{equation}
    M(x)=BT_{\rm BAO}(x')+A(x),
\end{equation}
where the position of the BAO feature is given in terms of the BAO-scaling parameter, $\alpha$. $x'$ is given by $\alpha\theta$, $\ell/\alpha$ and $s_\perp\alpha$ for ACF, APS and PCF, respectively. By definition,
\begin{equation}\label{eq:alpha}
    \alpha(\zeff)=\frac{D_M(\zeff)}{r_d}\left[\frac{D_M^{\rm ref}(\zeff)}{r_d^{\rm ref}}\right]^{-1},
\end{equation}
where $D_M(z)$ is the comoving angular diameter distance and $r_d$ is the sound horizon scale at recombination. ``ref'' refers to quantities evaluated at the reference cosmology (\planck).

\subsection{Systematics}

We mitigate the impact of observational systematics by applying correcting weights obtained with the \emph{Iterative Systematics Decontamination} method \cite{y3-galaxyclustering} to our galaxy sample (see \cite{mena2024dark} for details).

\subsection{Pre-Unblinding Tests}

The analysis and most of the paper writing were performed blind. Before unblinding, we require our data to pass a battery of tests. Some of these include whether we have a detection of the BAO feature in the combined fit and each redshift bin individually or not, if our measurement is robust against analysis choices, and if our three estimators give consistent results. All these tests pointed to the robustness of our measurement, so we unblinded (see \cite{BAOkey} for more details).

\section{Results}

\begin{figure}
    \centering
    \begin{minipage}{0.317\textwidth}
        \centering
        \includegraphics[width=\linewidth]{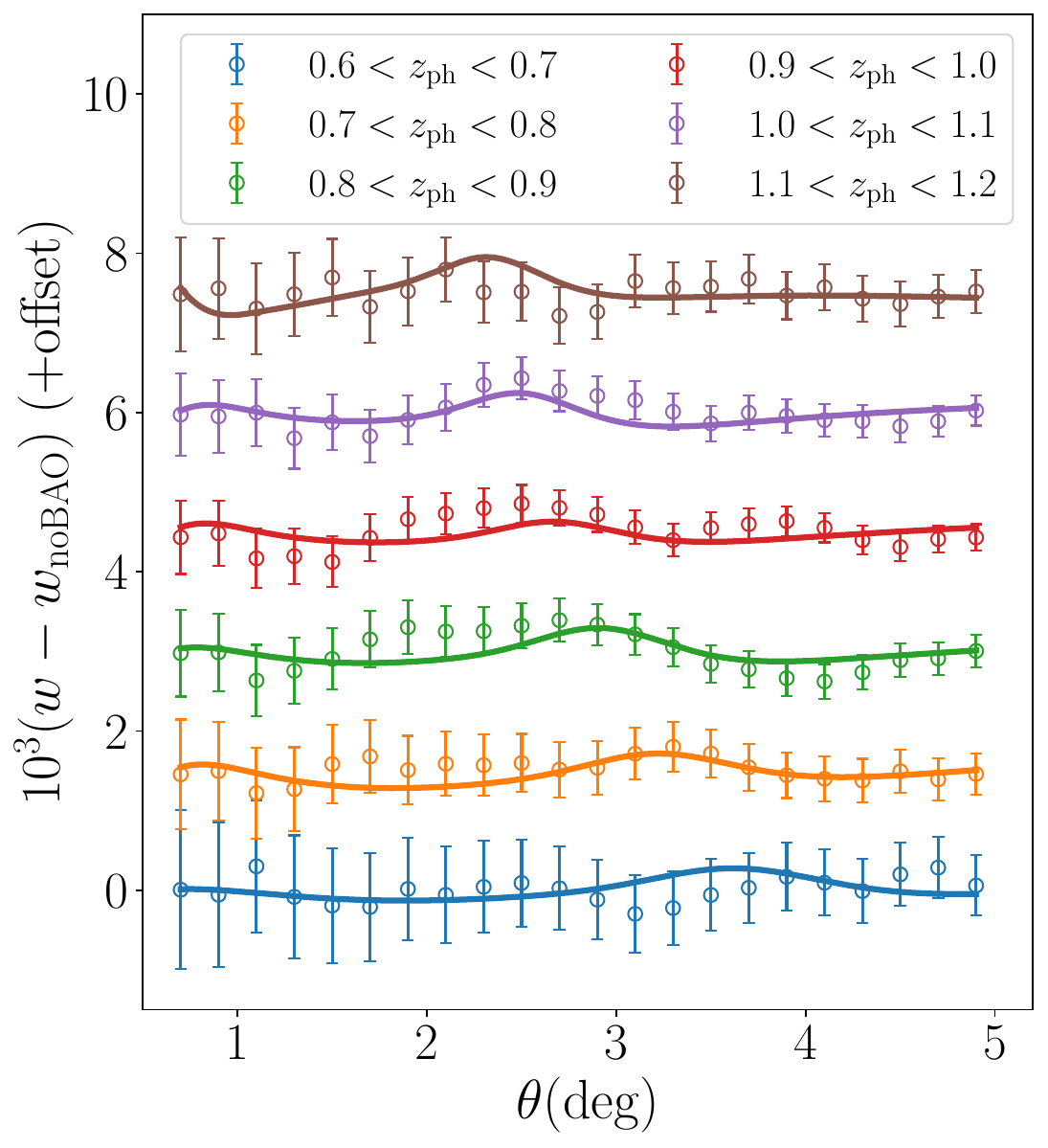}
    \end{minipage}
    \begin{minipage}{0.317\textwidth}
        \centering
        \includegraphics[width=\linewidth]{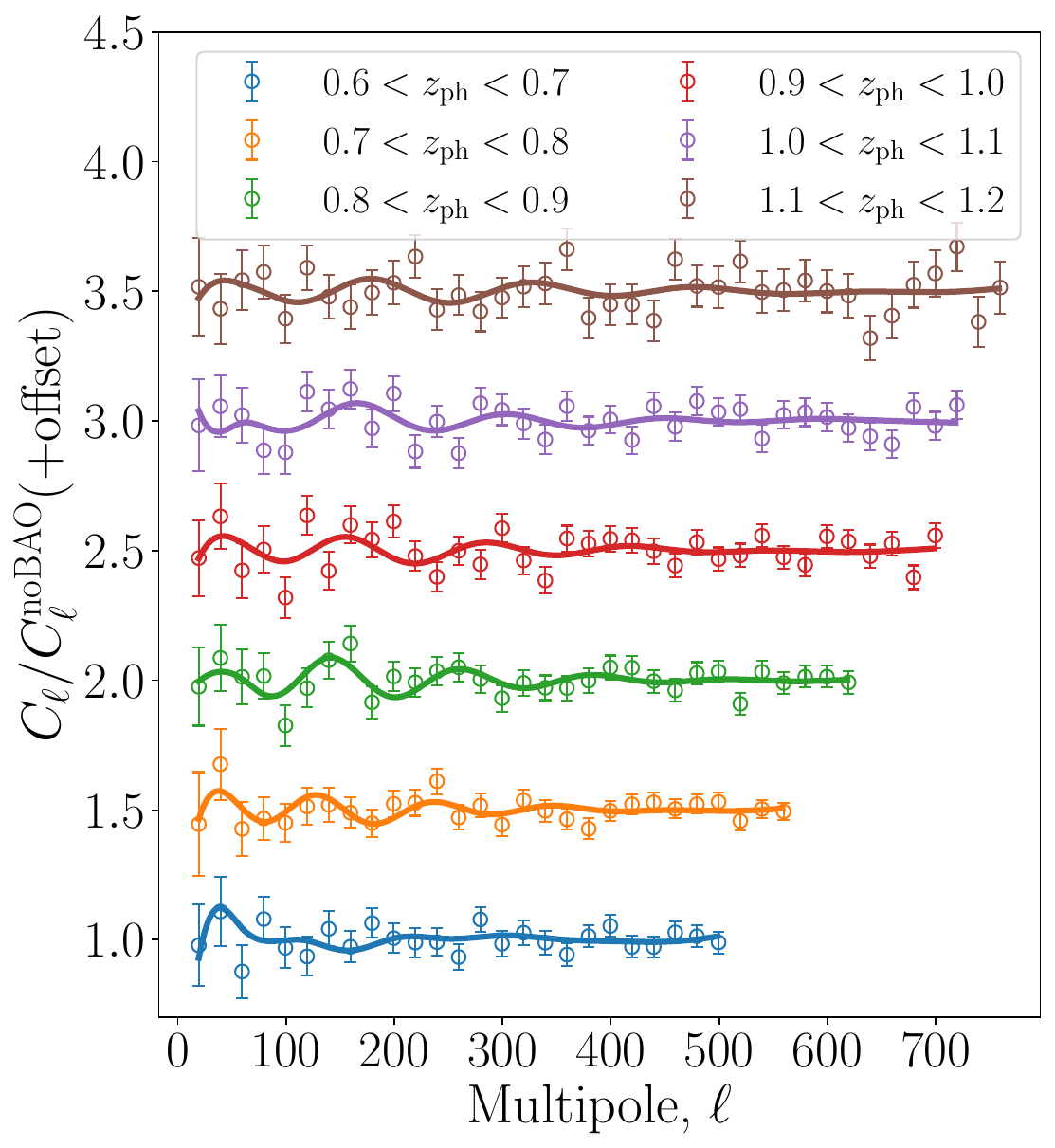}
    \end{minipage}
    \begin{minipage}{0.335\textwidth}
        \centering
        \includegraphics[width=\linewidth]{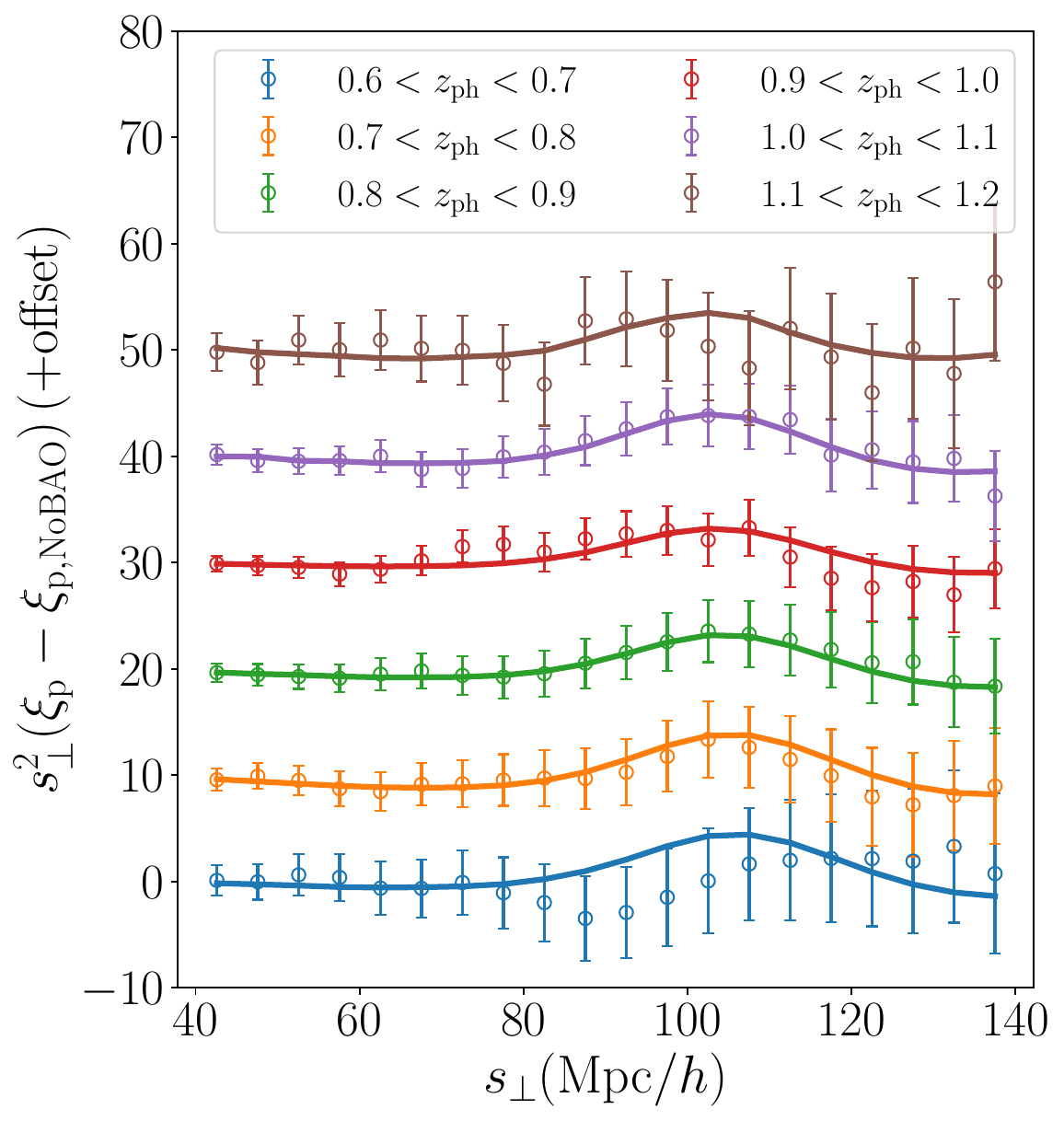}
    \end{minipage}
    \caption{Isolated BAO feature, measured using the ACF (left), APS (center) and PCF (right). Measurements are shown as markers with error bars, while the best-fit model is shown as solid lines. For each estimator, the fit is performed using the clustering signal of the 6 redshift bins together (accounting for their covariances).}
    \label{fig:BAO_signal}
\end{figure}

In Fig. \ref{fig:BAO_signal}, we show the clustering measurements for our different estimators and the best-fit model.
Our fiducial results are $\alpha=0.9517\pm0.0227$, $0.9617\pm0.0224$ and $0.9553\pm0.0201$ for ACF, APS and PCF, respectively. Our consensus measurement (combination of the three) is
\begin{eqnarray}
    \alpha(\zeff) = 0.9571 \pm 0.0196 & {\rm [ stat. ]},\\ 
    \hphantom{\alpha(\zeff) = 0.9571} \pm 0.0041 & {\rm [  sys. ]},\\
    \alpha(\zeff) = 0.9571 \pm 0.0201 & {\rm [  tot. ]},\label{eq:combined_result}
\end{eqnarray}
where the [sys.] contribution comes from modeling and redshift calibration, see \cite{BAOkey}. This result has a fractional error of 2.1\% (the smallest for a photometric survey ever), and is consistent with \planck ($\alpha=1$) at 2.13$\sigma$. It can also be expressed as $D_M(\zeff)/r_d = 19.51 \pm 0.41$.

\begin{figure}
    \centering
    \includegraphics[width=\linewidth]{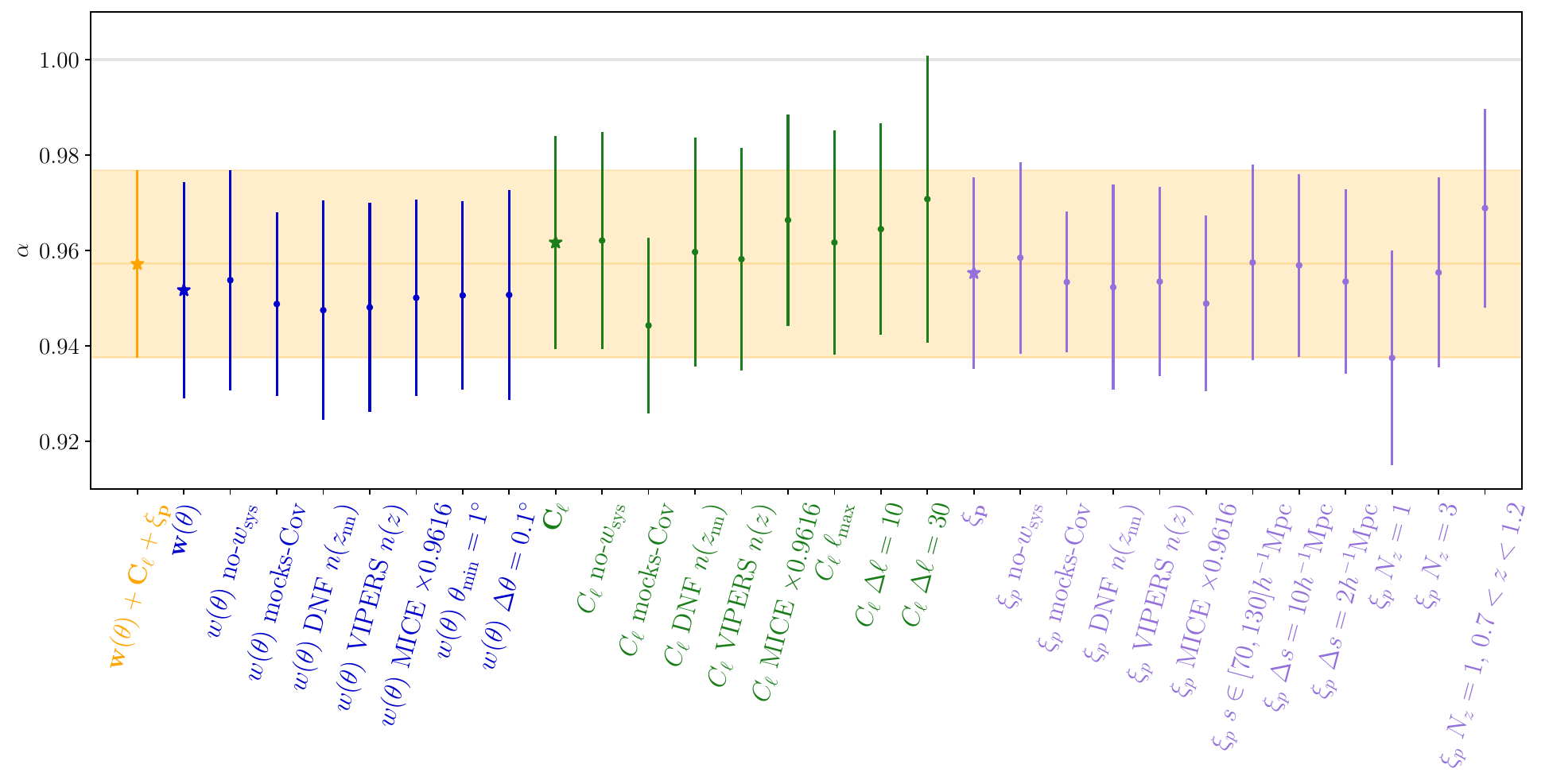}
    \caption{Main BAO measurement shown with an orange star and an orange shaded area, together with several variations of the analysis. Variations of the ACF, APS and PCF analyses are presented in blue, green and purple, respectively, with the fiducial settings being represented by the first entry for each of them.}
    \label{fig:robustness}
\end{figure}

\begin{figure}
    \centering
    \includegraphics[width=\linewidth]{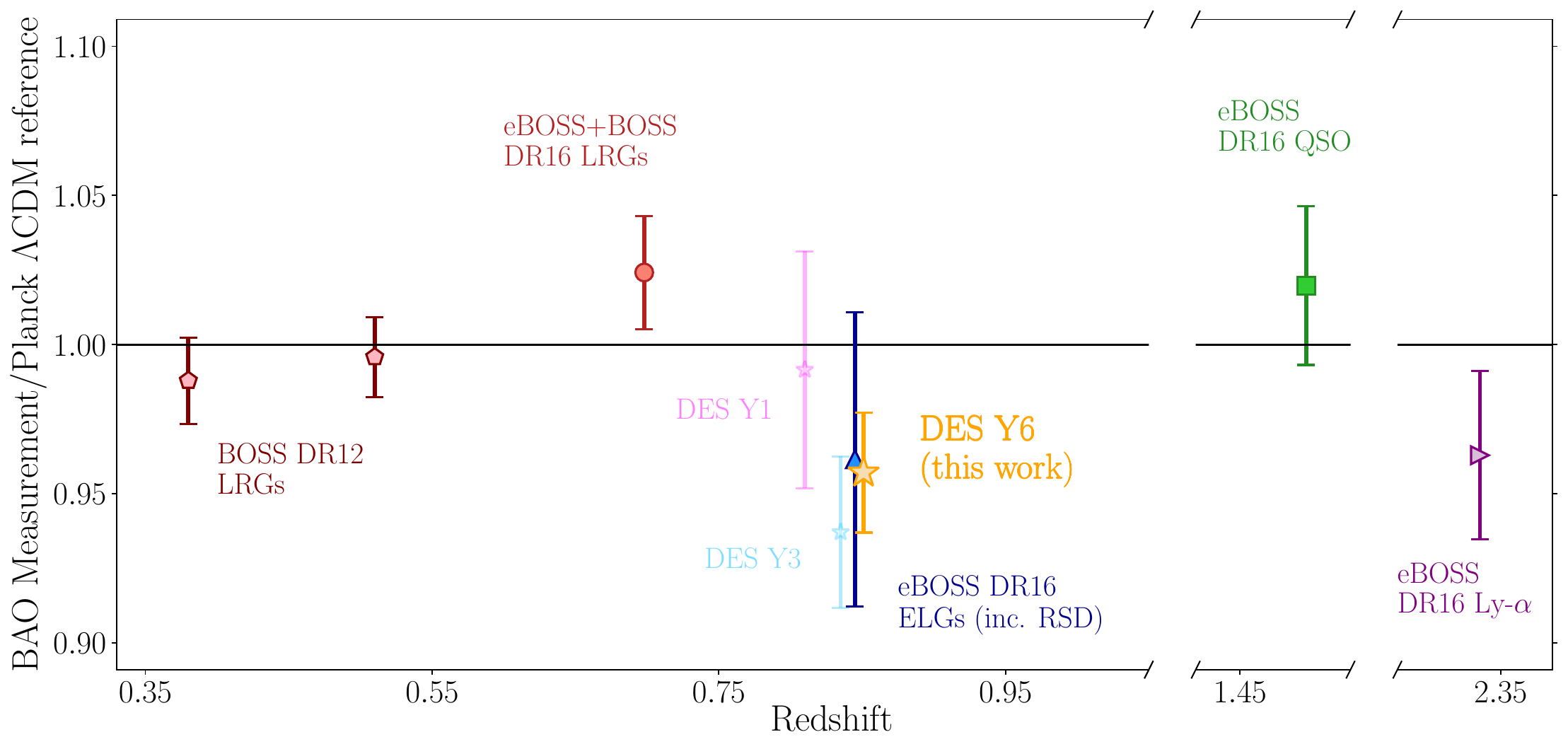}
    \caption{Ratio between the $D_M(z)/r_d$ measured using the BAO feature at different redshifts for several galaxy surveys and the prediction from the cosmological parameters determined by \planck, assuming $\Lambda$CDM. We include a series of measurements by SDSS, and also the DES Y1 and Y3 results. Our DES final data set measurement is shown as an orange star. This represents the most updated angular BAO distance ladder at the closure of Stage III.}
    \label{fig:ladder}
\end{figure}

In Fig. \ref{fig:robustness}, we show our main measurement, together with the results obtained with several variations of the analysis. We find that it is robust against variations in the fiducial settings, different clustering estimators and data calibration (systematics).
Finally, in Fig. \ref{fig:ladder} we show the angular BAO distance ladder at the end of Stage III. We find that our measurement is competitive with the results from the Sloan Digital Sky Survey (SDSS), and is the most precise one at $z>0.75$ at the end of Stage III.

\section*{References}

\bibliography{JuanMena}

\end{document}